\input{aipcheck}

\edef\optionlist{%
   \variorefoptionifavailable        
   draft,%
   \psnfssproblemoption              
   tnotealph}

\documentclass[\optionlist]{aipproc}

\newcommand{\be}{\begin{equation}}
\newcommand{\ee}{\end{equation}}
\newcommand{\bea}{\begin{eqnarray}}
\newcommand{\eea}{\end{eqnarray}}

\def\simg{{\ \lower-1.2pt\vbox{\hbox{\rlap{$>$}\lower6pt\vbox{\hbox{$\sim$}}}}\ }}
\def\siml{{\ \lower-1.2pt\vbox{\hbox{\rlap{$<$}\lower6pt\vbox{\hbox{$\sim$}}}}\ }}

\layoutstyle{6x9}

\usepackage{ttct0001}

\usepackage{shortvrb}
\MakeShortVerb\|

\makeatother

\begin{document}

\title
      [Color Superconductivity: Symmetries and Effective Lagrangians]
      {Color Superconductivity:\\ Symmetries and Effective Lagrangians}


\author{Francesco Sannino}{
  address={\centerline{{\rm NORDITA
  }} Blegdamsvej 17,Copenhagen \O, DK-2100, Denmark. },
  email={francesco.sannino@nbi.dk},
  homepage={http://www.nordita.dk/~sannino/}
  thanks={QCD&Work Conference BARI, Italy.}
}

\copyrightyear  {2001}

\begin{abstract}
I briefly review the symmetries and the associated low energy
effective Lagrangian for two light flavor Color Superconductivity
(2SC).
\end{abstract}

\date{\today}

\maketitle

\section{2SC Symmetries and Effective Lagrangian}
\label{review}


Quark matter at very high density  is expected to behave as a
color superconductor \cite{RW}.  Possible phenomenological
applications include the description of quark stars, neutron star
interiors, the physics near the core of collapsing stars and
supernova explosions \cite{RW,Hong:2001gt,OS}. The color
superconductive phase is characterized by its gap energy
($\Delta$) associated to quark-quark pairing which leads to the
spontaneous breaking of the color symmetry.

To describe low energy physical processes, where perturbation
theory is not applicable, effective Lagrangians based on the
global symmetries of the underlying theory are known to play a
relevant role. In the case of Color superconductivity effective
Lagrangians describe the interactions among the excitations near
the fermi surface. The three flavor case (CFL) has been developed
in \cite{CG}. The low-energy effective Lagrangian for the in
medium fermions and the broken sector of the $SU_c(3)$ color group
for 2SC has been constructed in Ref.~\cite{CDS}. The effective
theories encoding also the electroweak interactions for the
low-energy excitations in the 2SC and CFL case can be found in
\cite{CDS2001}. The light glueball Lagrangian of the unbroken
$SU_c(2)$ Yang-Mills sector of the 2SC phase has been constructed
in \cite{OSPLB}.

Here I summarize the effective low energy Lagrangian for two
flavors which contains all of the relevant degrees of freedom.
{}First I review the low-energy effective Lagrangian for the 2SC
phase of QCD \cite{CDS,CDS2001}. The latter describes the, in
medium, fermions and the broken $SU_c(3)$ gluon sector. I then
show how to build the effective Lagrangian describing the light
glueballs associated with the unbroken $SU_c(2)$ color subgroup by
using the information inherent to the trace anomaly and the medium
effects related to a non-vanishing dielectric constant first
presented in \cite{rischke2k} and confirmed within a different
formalism in \cite{Casalbuoni:2001ha}. Finally the, in medium,
glueball to two photon decay process is estimated. The present
talk is based on the papers \cite{CDS,CDS2001,OSPLB,S,HSaS}.

Quantum Chromo Dynamics with two flavors has gauge symmetry
$SU_{c}(3)$ and global symmetry
\begin{equation} SU_{L}(2)\times
SU_{R}(2)\times U_{V}(1)\ .\end{equation} \noindent  At high
matter density a color superconductive phase sets in and the
associated diquark condensate leaves invariant the following
symmetry group:\begin{equation} \left[ SU_{c}(2)\right] \times
SU_{L}(2)\times SU_{R}(2)\times \widetilde{U}_{V}(1) \
,\end{equation} where $\left[ SU_{c}(2)\right] $ is the unbroken
part of the gauge group. The $\widetilde{U}_{V}(1)$ generator
$\widetilde{B}$ is the following linear combination of the
previous $U_{V}(1)$ generator $B=\frac{1}{3}{\rm diag}(1,1,1)$ and
the broken diagonal generator of the $SU_{c}(3)$ gauge group
$T^{8}=\frac{1}{2\sqrt{3}}\,{\rm diag}(1,1,-2)$:
$\widetilde{B}=B-\frac{2\sqrt{3}}{3}T^{8} \label{residue}$. The
quarks with color $1$ and $2$ are neutral under $\widetilde{B}$
and consequently the condensate too ($\widetilde{B}$ is
$\sqrt{2}\widetilde{S}$ of Ref.~\cite{CDS}). The superconductive
phase for $N_{f}=2$ possesses the same global symmetry group of
the confined Wigner-Weyl phase \cite{S}. In Reference \cite{S}, it
was shown that the low-energy spectrum, at finite density,
displays the correct quantum numbers to saturate the 't~Hooft
global anomalies \cite{tHooft}. It was also observed that QCD at
finite density can be envisioned, from a global symmetry and
anomaly point of view, as a chiral gauge theory. In Reference
\cite{HSaS} it was then seen, by using a variety of field
theoretical tools, that global anomaly matching conditions hold
for any cold but dense gauge theory.

The lowest lying excitations are protected from acquiring a mass
by the aforementioned constrains and dominate the low-energy
physical processes. The low-energy theorems governing their
interactions can be usefully encoded in effective Lagrangians. The
dynamics of the Goldstone bosons is efficiently encoded in a
non-linear realization framework. Here, see \cite{CDS}, the
relevant coset space is $G/H$ with $G=SU_{c}(3)\times U_{V}(1)$
and $H=SU_{c}(2)\times \widetilde{U}_{V}(1)$ is parameterized by
\begin{equation}{\cal V}=\exp (i\xi ^{i}X^{i})\ ,\end{equation}
where $\{X^{i}\}$ $i=1,\cdots ,5$ belong to the coset space $G/H$
and are taken to be $X^{i}=T^{i+3}$ for $i=1,\cdots ,4$ while
$X^{5}=B+\frac{\sqrt{3}}{3}T^{8}={\rm
diag}(\frac{1}{2},\frac{1}{2},0) \label{broken}$. $T^{a}$ are the
standard generators of $SU(3)$. The coordinates
\begin{equation}
\xi ^{i}=\frac{\Pi ^{i}}{f}\quad i=1,2,3,4\ ,\qquad \xi
^{5}=\frac{\Pi ^{5}}{\widetilde{f}}\ ,
\end{equation}
via $\Pi $ describe the Goldstone bosons.

${\cal V}$ transforms non linearly \begin{equation}{\cal V}(\xi
)\rightarrow u_{V}\,g{\cal V}(\xi )h^{\dagger }(\xi
,g,u)h_{\widetilde{V}}^{\dagger }(\xi ,g,u)\ ,
\label{nl2}\end{equation} with $u_{V}\in U_{V}(1)$, $g\in
SU_{c}(3)$, $h(\xi ,g,u)\in SU_{c}(2)$ and $h_{\widetilde{V}}(\xi
,g,u)\in \widetilde{U}_{V}(1)$. It is, also, convenient to define:
\begin{equation}
\omega _{\mu }=i{\cal V}^{\dagger }D_{\mu }{\cal V}\quad {\rm
with}\quad D_{\mu }{\cal V}=(\partial _{\mu }-ig_{s}G_{\mu }){\cal
V}\ ,
\end{equation}
with gluon fields $G_{\mu }=G_{\mu }^{m}T^{m}$.
Following \cite{CDS} we decompose $\omega _{\mu }$ into
\begin{equation}
\omega _{\mu }^{\parallel }=2S^{a}{\rm Tr}\left[ S^{a}\omega _{\mu
}\right] \quad {\rm and}\quad \omega _{\mu }^{\perp }=2X^{i}{\rm
Tr}\left[ X^{i}\omega _{\mu }\right] \ ,
\end{equation}
where $S^{a}$ are the unbroken generators of $H$ with
$S^{1,2,3}=T^{1,2,3}$, $S^{4}=\widetilde{B}\,/\sqrt{2}$. Summation
over repeated indices is assumed.

To be able to include the in medium fermions in the picture we
define:
\begin{equation}
\widetilde{\psi}={\cal V}^{\dagger }\psi \ ,  \label{mq}
\end{equation}
transforming as $\widetilde{\psi}\rightarrow
h_{\widetilde{V}}(\xi,g,u)h(\xi ,g,u)\widetilde{ \psi}$ and $\psi$
possesses an ordinary quark transformations (as Dirac spinor).

The simplest non-linearly realized effective Lagrangian describing
in medium fermions, the five gluons and their self interactions,
up to two derivatives and quadratic in the fermion fields is:
\begin{eqnarray}
{\cal L}=~ &&f^{2}a_{1}{\rm Tr}\left[ \,\omega _{0}^{\perp }\omega
_{0}^{\perp }-{\alpha }_{1}\vec{\omega}^{\perp
}\vec{\omega}^{\perp }\, \right] + f^{2}a_{2}\left[ {\rm Tr}\left[
\,\omega _{0}^{\perp }\,\right] {\rm Tr}\left[ \,\omega
_{0}^{\perp }\,\right] -{\alpha }_{2}{\rm Tr}\left[
\,\vec{\omega}^{\perp }\,\right] {\rm Tr}\left[
\,\vec{\omega}^{\perp }\, \right] \right]  \nonumber \\ &+&
b_{1}\overline{\widetilde{\psi }}i\left[ \gamma ^{0}(\partial
_{0}-i\omega _{0}^{\parallel })+\beta _{1}\vec{\gamma}\cdot \left(
\vec{ \nabla}-i\vec{\omega}^{\parallel }\right) \right]
\widetilde{\psi } + b_{2} \overline{\widetilde{\psi }}\left[
\gamma ^{0}\omega _{0}^{\perp }+\beta _{2} \vec{\gamma}\cdot
\vec{\omega}^{\perp }\right] \widetilde{\psi } \nonumber \\
&+&m_{M}\overline{\widetilde{\psi }^{C}}\gamma
^{5}(iT^{2})\widetilde{\psi }+ {\rm h.c.}\ , \label{cadusa}
\end{eqnarray}
where $\widetilde{\psi }^{C}=i\gamma ^{2}\widetilde{\psi }^{\ast
}$, $i,j=1,2 $ are flavor indices and \begin{equation}
T^{2}=S^{2}=\frac{1}{2}\left(
\begin{array}{ll}
\sigma ^{2} & 0 \\ 0 & 0
\end{array}
\right)\ , \end{equation} $a_{1},~a_{2},~b_{1}$ and $b_{2}$ are
real coefficients while $m_{M}$ is complex. The breaking of
Lorentz invariance to the $O(3)$ subgroup, following \cite{CG},
has been taken into account by providing different coefficients to
the temporal and spatial indices of the Lagrangian, and it is
encoded in the coefficients $\alpha $s and $\beta $s. For
simplicity, the flavor indices are omitted. {}From the last two
terms, representing a Majorana mass term for the quarks, we deduce
that the massless degrees of freedom are the $\psi _{a=3,i}$ which
possess the correct quantum numbers to match the 't~Hooft anomaly
conditions \cite{S}. The generalization to the electroweak
processes relevant for the cooling history of compact stars has
been investigated in \cite{CDS2001}.


\section{The $SU_c(2)$ Glueball Effective Lagrangian}
\label{Glueball}

The $SU_c(2)$ gauge symmetry does not break spontaneously and it
is expected to confine. If the new confining scale is lighter than
the superconductive quark-quark gap the associated confined
degrees of freedom (light glueballs) \cite{OSPLB} can play,
together with the true massless quarks a relevant role for the
physics of Quark Stars featuring a 2SC superconductive surface
layer \cite{OS}.

\noindent Indeed, according to the findings in \cite{rischke2k},
the medium does lead to partial $SU_c(2)$ screening. In other
words the medium is polarizable, i.e., acquires a dielectric
constant $\epsilon$ different from unity (in fact $\epsilon \gg 1$
in the 2SC case \cite{rischke2k}) leading to an effectively
reduced gauge coupling constant. By assuming locality the
$SU_c(2)$ effective action takes the form \cite{rischke2k}:
\begin{equation}
S_{eff}=\int \, d^4x
\left[\frac{\epsilon}{2}{\vec{E}^a}\cdot{\vec{E}^a}-\frac{1}{2\lambda}\vec{B}^a\cdot
\vec{B}^a\right] \label{sefu2}
\end{equation}
with $a=1,2,3$ and $E_{i}^a \equiv F^a_{0i}$ and  $B^a_{i}\equiv
\frac{1}{2}\epsilon_{ijk} F^a_{jk}$. Here one assumes an expansion
in powers of the fields and derivatives. The gluon speed in this
regime is $v=1/\sqrt{\epsilon \lambda}$. \noindent In Reference
\cite{rischke2k} the $\epsilon$ and $\lambda$ were obtained:
\begin{equation}
\epsilon =1 + \frac{g_s^2 \mu^2}{18 \pi^2 \Delta^2}\ , \qquad
\lambda =1 \ . \label{el}
\end{equation}
Equation (\ref{el}) than suggests that a 2SC color superconductor
can have a large positive dielectric constant. This implies that
the Coulomb potential between $SU_c(2)$ color charges is reduced
in the 2SC medium. $SU_c(2)$ glueballs like particles are expected
to emerge. These particles are light with respect to $\Delta$. So,
the low-energy $SU_c(2)$ theory should be well represented by the
effective Lagrangian describing its hadronic low lying states.
This Lagrangian has to be added to the one of Eq.~(\ref{cadusa})
\cite{CDS} and it has been constructed in \cite{OSPLB}.

We first rescale the coordinates and the $SU_c(2)$ fields as
follows:
\begin{eqnarray}
\hat{x}^0&=&\frac{x^0}{\sqrt{\lambda \epsilon}} \ , \qquad \hat{g}
=
g_s\left(\frac{\lambda}{\epsilon}\right)^{\frac{1}{4}} \qquad
\hat{A}_0^a = \lambda^{\frac{1}{4}}\epsilon^{\frac{3}{4}}A_0^a \ ,
\qquad \hat{A}_i^a =
\lambda^{-\frac{1}{4}}\epsilon^{\frac{1}{4}}A_i^a \ .
\label{drescaling}
\end{eqnarray}
The $SU_c(2)$ action now becomes: \begin{equation}S_{SU(2)} =
-\frac{1}{2} \int \, d^4 \hat{x}\, {\rm Tr} \left[ \hat{F}_{\mu
\nu} \hat{F}^{\mu \nu} \right]\label{effsu2}\ ,\end{equation} and
$\hat{F}_{\mu,\nu}=\hat{\partial}_{\mu}\hat{A}_{\nu} -
\hat{\partial}_{\nu} \hat{A}_{\mu} +
i\,\hat{g}\left[\hat{A}_{\mu},\hat{A}_{\nu}\right]$ with
$\hat{A}_{\mu}=\hat{A}^a_{\mu}T^a$ and $a=1,2,3$. \noindent The
low-energy effective 3 gluon dynamics in the color superconductor
medium (with non-vanishing dielectric constant and magnetic
permeability) is similar to the in vacuum theory. The expansion
parameter is: $\displaystyle{
\hat{\alpha}=\frac{\hat{g}^2}{4\pi}=\frac{g_s^2}{4\pi}
\sqrt{\frac{\lambda}{\epsilon}}}$. Notice that $g_s$ is the
$SU_c(3)$ coupling constant evaluated at the scale $\mu$ while we
now, following Ref.~\cite{rischke2k}, interpret $\hat{g}$ as the
$SU_c(2)$ coupling at $\Delta$. The matching of the scales is
encoded in $\sqrt{\lambda/\epsilon}$.

The, in medium, anomaly-induced effective Lagrangian is based on
the trace anomaly arising from the rescaled $SU_c(2)$ \cite{SS}:
\begin{equation}
\hat{\theta}_{\mu}^{\mu}=-\frac{\beta(\hat{g})}
{2\hat{g}}\hat{F}^{\mu\nu}_a\,\hat{F}_{\mu\nu;a}\equiv
\frac{2b}{v}\, H \ ,\label{trace}
\end{equation}
with $a=1,2,3$ and we have defined $\beta(\hat{g}) = -b {\hat
{g}^3}/16 \pi^2 $. At one loop $b=\frac{11}{3}N_c$ with $N_c=2$
the color number. $H$ is the composite field describing, upon
quantization, the scalar glueball \cite{schechter} in medium and
possesses mass-scale dimensions 4. The specific velocity
dependence is introduced to properly account for the velocity
factors.

The complete simplest light glueball action in the unrescaled
coordinates for the, in medium, Yang-Mill theory is:
\begin{eqnarray}
S_{G-ball}=\int
&d^4x&\left\{\frac{c}{2}\sqrt{b}\,H^{-\frac{3}{2}}\left[\partial^{0}
H
\partial^{0}H - v^2
\partial^iH
\partial^iH\right]   -\frac{b}{2}
H\log\left[\frac{H}{\hat{\Lambda}^4}\right] \right\} \ .
\label{G-ball}
\end{eqnarray}
The glueballs move with the same velocity $v$ as the underlying
gluons in the 2SC color superconductor. $\hat{\Lambda}$ is the
intrinsic scale associated with the theory and can be less than or
of the order of few MeVs \cite{rischke2k,OSPLB} while $c$ is a
constant of order unity.

\noindent The glueballs are light (with respect to the gap) and
might barely interact with the ungapped fermions. They are stable
with respect to the strong interactions unlike ordinary glueballs.
We define the mass-dimension one glueball field $h$ via
\begin{equation} H=\langle H\rangle e^{\frac{h}{F_h}} \ . \end{equation} By requiring a
canonically normalized kinetic term for $h$ one finds
$F_h^2=\frac{c}{\sqrt{2}} \sqrt{2b\langle H\rangle}$, while the
glueball mass term is $M^2_h=\frac{\sqrt{b}}{2c}\sqrt{\langle
H\rangle}=\frac{\sqrt{b}}{2c\sqrt{e}} \hat{\Lambda}^2$, which is
clearly of the order of $\hat{\Lambda}$ since $c$ is a positive
constant of order unity.

Once created, the light $SU_c(2)$ glueballs are stable against
strong interactions but not with respect to electromagnetic
processes. Indeed, the glueballs couple to two photons via virtual
quark loops.

The relevant Lagrangian term, at non zero baryon density, obtained
by saturating the electromagnetic trace anomaly is \cite{OSPLB}:
\begin{equation} {\cal
L}_{h\gamma\gamma}=\frac{\widetilde{e}^2}{48\pi^2}
\frac{M_h}{\sqrt{2b\langle H  \rangle}} \left[\sum_{quarks}
\widetilde{Q}_{quarks}^2\right]
h\,\widetilde{F}_{\mu\nu}\widetilde{F}^{\mu \nu} \ ,
\end{equation}
with $\widetilde{F}_{\mu\nu}=\partial_{\mu}\widetilde{A}_{\nu}-
\partial_{\nu}\widetilde{A}_{\mu}$.
Here $\widetilde{A}_{\mu}$ is the in medium photon field
corresponding to the following massless linear combination of the
old photon and the eighth gluon \cite{charges,CDS2001}:
\begin{equation}
\widetilde{A}_{\mu }=\cos \theta _{Q}A_{\mu }-\sin \theta
_{Q}G_{\mu }^{8}\ ,
\end{equation}
with $\tan\theta _{Q}=e/(\sqrt{3}g_s)$. The new electric constant
is related to the in vacuum one via $\widetilde{e}=e\,\cos \theta
_{Q}$. $\widetilde{Q}$ is the new electric charge operator
associated with the field $\widetilde{A}_{\mu }$ with
$\widetilde{Q}={\tau ^{3}\times {\mbox{\bf
1}}+\frac{\widetilde{B}-L}{2}}=Q\times {\mbox{\bf
1}}-\frac{1}{\sqrt{3}}{\mbox{\bf 1}\times }T^{8}
\label{newcharge}$, where $L=0$ is the lepton number, $\tau^3$ the
standard Pauli's matrix, $Q$ the quark matrix, while the new
baryon number is $\widetilde{B}$, and following the notation of
Ref.~\cite{CDS2001} we have ${\rm flavor}_{2\times 2}\times {\rm
color}_{3\times 3}$. This leads to the following decay width of
the glueballs into two photons in medium:
\begin{eqnarray}
 \Gamma\left[h\rightarrow
\gamma\gamma\right] \approx 1.2\times 10^{-2} \cos\theta_{Q}^4
\left[\frac{M_h}{1~{\rm MeV}}\right]^5~{\rm eV} \ ,
\end{eqnarray}
where $\alpha=e^2/4\pi \simeq 1/137$. {}For illustration purposes
we consider a glueball mass of the order of $1$~MeV which leads to
a decay time $\tau\sim~5.5\times~10^{-14}s$. We used
$\cos\theta_Q\sim~1$ since $\theta_Q \sim 2.5^{\circ}$
\cite{OSPLB}. While we are aware of the possible contribution from
other hadrons to the saturation of the electromagnetic trace
anomaly \cite{GJJS,GJJS2}, here we assume it to be dominated by
the $SU_c(2)$ glueballs. In any case, it is hard to imagine the
photon decay process to be completely switched-off. This shows
that a consistent portion of the glue ($3/8$ or $37.5\%$) filling
the 2SC medium is very rapidly and efficiently converted into
electromagnetic radiation.

\section{Conclusion}
\label{conclusion}
 I reviewed the symmetries and the low energy
effective Lagrangian for two flavor Color Superconductivity. The
effective Lagrangian describes the, in medium, fermions and the
broken $SU_c(3)$ gluon sector. The theory has then been extended
to incorporate the relevant confining and light (with respect to
the gap), $SU_c(2)$ degrees of freedom, i.e. glueballs. It is
shown that the light glueballs are unstable to photon decay and
estimated the, in medium, two photon decay rate. The present
analysis is limited to the zero temperature and high matter
density case. However it might be relevant to investigate the role
played by a non zero temperature \cite{MS}.

\begin{theacknowledgments}

I thank Roberto Casalbuoni, Zhiyong Duan, Stephen D.~Hsu, Rachid
Ouyed and Myck Schwetz for sharing part of the work on which this
talk is based. For discussions and careful reading of the
manuscript I thank Nils Marchal while I am indebted to Joseph
Schechter for enlightening discussions and continuous
encouragement.
\end{theacknowledgments}


\begin{thebibliography}{99}



\bibitem{RW} For reviews see K.~Rajagopal, F.~Wilczek
hep-ph/0011333;
M.~Alford,
hep-ph/0102047
;
S.~D.~Hsu,
hep-ph/0003140
, and references therein.

\bibitem{Hong:2001gt}
D.~K.~Hong, S.~D.~Hsu and F.~Sannino,
hep-ph/0107017.

\bibitem{OS} R.~Ouyed and F.~Sannino, astro-ph/0103022.


\bibitem{CG}  R.~Casalbuoni and R.~Gatto, Phys.~Lett.~B{\bf 464}, 11 (1999);
Phys.~Lett.~B{\bf 469}, 213 (1999).

\bibitem{CDS}  R.~Casalbuoni, Z.~Duan and F.~Sannino, Phys.~Rev.~D{\bf 62},
094004, (2000).



\bibitem{CDS2001} R.~Casalbuoni, Z.~Duan and F.~Sannino,
hep-ph/0011394. Phys.~Rev.~D{\bf 63}, 114026, (2001).

\bibitem{OSPLB} R.~Ouyed and F.~Sannino, Phys.~Lett.~B{\bf 511}, 66, (2001).


\bibitem{rischke2k}  D.H. Rischke, D.T. Son, M.A. Stephanov,
hep-ph/0011379.


\bibitem{Casalbuoni:2001ha}
R.~Casalbuoni, R.~Gatto, M.~Mannarelli and G.~Nardulli,
hep-ph/0107024.

\bibitem{S}  F.~Sannino, Phys.~Lett.~B{\bf 480}, 280, (2000).


\bibitem{HSaS}  S.~Hsu, F.~Sannino and M.~Schwetz, hep-ph/0006059.

\bibitem{tHooft}  G.~'t~Hooft, in: Recent Developments in Gauge Theories,
eds., G.~'t~Hooft (Plenum Press, New York, 1980).

\bibitem{SS} F.~Sannino and J.~Schechter, Phys.~Rev.~D{\bf 60},
056004, (1999).




\bibitem{schechter}
J.~Schechter, Phys. Rev. {\bf D21}, 3393 (1980).


\bibitem{charges} M.~Alford,~J.~Berges~and~K.~Rajagopal, Nucl. Phys. B{\bf 571}, 269 (2000).



\bibitem{GJJS}
H.~Gomm, P.~Jain, R.~Johnson and J.~Schechter, Phys. Rev. {\bf
D33}, 801 (1986).

\bibitem{GJJS2}
H.~Gomm, P.~Jain, R.~Johnson and J.~Schechter, Phys. Rev. {\bf
D33}, 3476 (1986).



\bibitem{MS} N.~Marchal and F.~Sannino, work in progress.

\end{thebibliography}
\end{document}